\documentclass[12pt]{article}   

\usepackage{geometry}                
\usepackage{graphicx}
\usepackage{amssymb}
\usepackage{epstopdf}
\usepackage{amsmath}
\usepackage{alltt}
\usepackage{wrapfig}
\usepackage{multirow}
\usepackage{rotating}

\begin{document}

\begin{LARGE}
\begin{center}
Phonon Quasidiffusion in Cryogenic Dark Matter Search Large Germanium Detectors
\end{center}
\end{LARGE}
\vspace{1 cm} 

\begin{large}
\begin{center}
S.W.~Leman$^{a*}$, B.~Cabrera$^b$, K.A.~McCarthy$^a$, M.~Pyle$^b$, R.~Resch$^c$, B.~Sadoulet$^d$, K.M.~Sundqvist$^d$, P.L.~Brink$^b$, M.~Cherry$^b$, E.~Do~Couto~E~Silva$^c$, E.~Figueroa-Feliciano$^a$, N.~Mirabolfathi$^d$, B.~Serfass$^d$, and A.~Tomada$^b$ 
\end{center}
\end{large}

\begin{small}
\begin{center}
(a) MIT Kavli Institute for Astrophysics and Space Research, Cambridge, MA, U.S.A. \\
(b) Department of Physics, Stanford University, Stanford, CA, U.S.A. \\
(c) Stanford Linear Accelerator Center, Menlo Park, CA 94309, U.S.A. \\
(d) Department of Physics, The University of California at Berkeley, Berkeley, CA, U.S.A. \\
* Corresponding Author's e-mail address: swleman@mit.edu \\
\end{center}
\end{small}

\begin{abstract}
We present results on quasidiffusion studies in large, 3~inch diameter, 1~inch thick [100] high purity germanium crystals, cooled to 50~mK in the vacuum of a dilution refrigerator, and exposed with 59.5~keV gamma-rays from an Am-241 calibration source. We compare data obtained in two different detector types, with different phonon sensor area coverage, with results from a Monte Carlo. The Monte Carlo includes phonon quasidiffusion and the generation of phonons created by charge carriers as they are drifted across the detector by ionization readout channels.

PACS numbers: 72.10.Di, 85.25.Oj, 95.35.+d

\end{abstract}

\section{Introduction}

The Cryogenic Dark Matter Search~\cite{Ahmed2009} utilizes silicon and germanium detectors to search for Weakly Interacting Massive Particle (WIMP) dark matter~\cite{Spergel2007,Tegmark2004} candidates. The silicon or germanium nuclei provide a target mass for WIMP-nucleon interactions. Simultaneous measurement of both phonon energy and ionization energy provide a powerful discriminator between electron-recoil (relatively high ionization) interactions and nuclear-recoil interactions (relatively low ionization). Background radiation primarily interacts through electron-recoils whereas a WIMP signal would interact through nuclear-recoils. The experiment is located in the Soudan Mine, MN, U.S.A.

The most recent phase of the CDMS experiment has involved fabrication, testing and commissioning of large, 3~inch diameter, 1~inch thick [100] germanium crystals. We present results on quasidiffusion studies in which the crystal was exposed with 59.5~keV gamma-rays from a collimated Am-241 calibration source. Prompt phonons are generated from electron-recoil interactions along with Luke-Neganov~\cite{Luke1988} phonons created by charges as they drift through the crystal via the ionization channels' electric field. In our Monte Carlo, phonon transport is described by quasidiffusion~\cite{Marris1990, Tamura1993}, which includes anisotropic propagation, isotope scattering and anharmonic decay, until the phonons are absorbed in either the Transition Edge Sensor (TES)~\cite{Irwin1995} based phonon channels (photolithographically patterned on the flat crystal surfaces) or lost in surface interactions. The small fraction of surface covered by the TES surfaces and the low surface losses result in phonon pulse constants much longer than the TES time constants allowing for good observation of the underlying phonon physics.

The phonon quasidiffusion measurements presented here differ from similar heat-pulse experiments in which the detector is partially submerged in a pumped liquid helium bath~\cite{Msall1997}. These crystals are situated in vacuum, which allows for reduced phonon losses at the surfaces. Additionally, these crystals were cooled to 50~mK in a dilution refrigerator. Phonon losses in the TESs and non-instrumented surfaces can be determined via the partitioning of energy in the phonon channels, and phonon propagation parameters are studied via signal timing in the phonon channels.

\section{CDMS Detectors}

The CDMS detectors are made of [001] high purity germanium crystals with [110] oriented flats. The CDMS detectors are 3~inches in diameter and 1~inch thick with a total mass of about 607~grams and come in two styles, the mZip and a prototype iZip style. 
		
\begin{figure}
\begin{tabular}{c}
\begin{minipage}{0.5\hsize}
\begin{center}
\includegraphics[width=7cm]{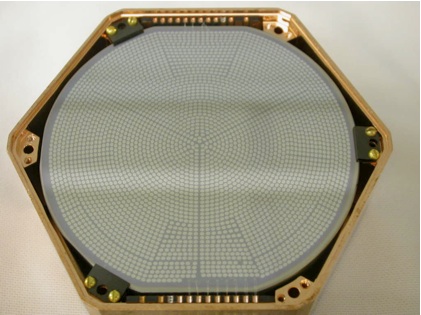}
\end{center}
\end{minipage}
\begin{minipage}{0.45\hsize}
\begin{center}
\includegraphics[width=7cm]{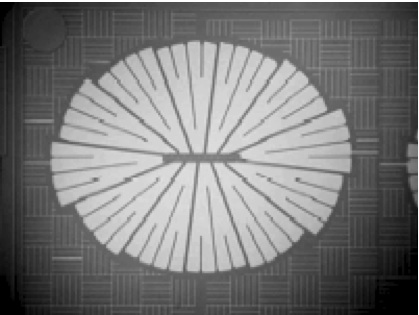}
\end{center}
\end{minipage}
\end{tabular}
\caption[] { \label{fig:mZip}(left) A CDMS ``mZip'' detector with photolithographically defined phonon sensors. The crystal is 3~inches in diameter and mounted in its copper housing. The detector is normally run with three inner channels and an outer, annular guard channel. Close inspection shows six pairs of wirebonds as the outer channel was segmented into three, individually instrumented channels.}
\caption[] { \label{fig:mZipStadium}(right) A close-up view of the aluminumÐtungsten phonon sensors. Phonons interact in the pie shaped aluminum fins where they create quasiparticles that diffuse into the rectangular shaped TES at the center of the ``pie''. The parquet pattern is non-instrumented aluminum which maintains a constant potential across this detector surface.}
\end{figure}

One side of the mZip detector has two concentric, photolithographically defined ionization channels. The outer ionization channel functions as a guard ring and is used to veto events located near the radial surface. The total aluminum coverage on this surface is $\sim$10\%, and phonons absorbed in this surface are not measured. The other side of the detector has four lithographically defined phonon channels as shown in Figure~\ref{fig:mZip} and~\ref{fig:mZipStadium}. Three of the channels are arranged in equal pie shaped wedges and the fourth channel is circular, at large radius, and also aids in determining the eventÕs radial location. All four of the channels have equal area coverage. The aluminum coverage on the phonon-channel side is 37\% active and 6.1\% passive.

\begin{figure}
\begin{tabular}{c}
\begin{minipage}{0.5\hsize}
\begin{center}
\includegraphics[width=7cm]{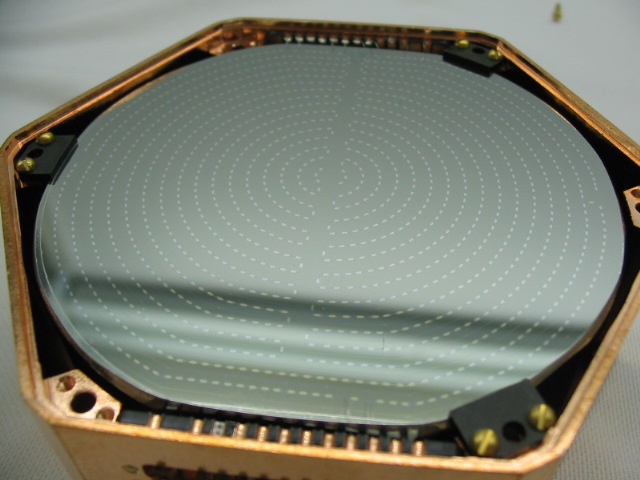}
\end{center}
\end{minipage}
\begin{minipage}{0.45\hsize}
\begin{center}
\includegraphics[width=7cm]{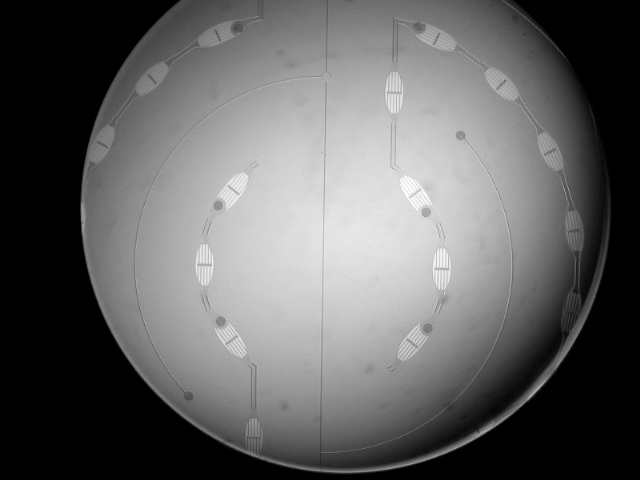}
\end{center}
\end{minipage}
\end{tabular}
\caption[] { \label{fig:iZip}(left) A CDMS ``iZip'' detector with photolithographically defined phonon sensors. The crystal is 3~inches in diameter and mounted in its copper housing. The top surface contains an outer, guard phonon sensor and two inner phonon sensors from which an eventÕs $x$ position estimate can be made. The opposite face (not shown) has a similar channel design, but rotated 90~degrees to determine an eventÕs $y$ position estimate.}
\caption[] { \label{fig:iZipTES}(right) Close-up view of the iZip phonon channel and ionization channel (thin lines in between the phonon sensors). The phonon channel is held at ground and the ionization channel is held at $\sim+(-)$2~V for the top (bottom) surfaces.}
\end{figure}

The iZip detector utilizes both anode and cathode lines on the same side of the detector similar to a Micro-Strip Gas Chamber (MSGC)~\cite{Knoll2000} as shown in Figure~\ref{fig:iZip} and~\ref{fig:iZipTES}. Unlike an MSGC however, these is a set of anode and cathode lines on both sides of the detector. This ionization channel design is used to veto events interacting near the detector surfaces. The phonon sensor design incorporates a guard ring channel, two channels on the top surface, from which $x$ position estimates can be made and two channels on the bottom surface, from which $y$ position estimates can be made. The total iZip aluminum coverage is, compared to the mZip, reduced significantly to 4.8\% active and 1.5\% passive per side.

Phonons are detected when they are absorbed into the aluminum fins, break Cooper pairs, and these Cooper pairs diffuse into the tungsten Transition Edge Sensors (TES). The TESs are superconductors, voltage biased within the superconducting to normal transition and the introduction of excited Cooper pairs raises their temperature which raises their resistance. Properly instrumenting the TESs allows the phonon flux power to be estimated.

\section{The CDMS Detector Monte Carlo, Phonons}

The CDMS detector Monte Carlo (CDMS-DMC) models both the charge and phonon dynamics; these signals combined provide information about event location within the detector, the amount of energy deposited in the detector and the interaction recoil type (electron-recoil vs. nuclear recoil). Partitioning of energy between the inner ionization channel and the outer, guard ionization channel combined with partitioning of energy between the four phonon channels provides information about event location. Event location provides a discriminator against surface event contamination, a class of event vetoed in low-background dark matter search operation. The ratio of energy in the ionization channel vs. energy in the phonon channel indicates recoil type and provides a discriminator against gamma and beta background radiation, additional event types vetoed in low-background operation. If and when a population of WIMPs is detected, the spectrum of deposited energy will provide information regarding the WIMP mass~\cite{Yellin2002}. A proper modeling of the CDMS detectors is therefore necessary to understand and interpret the detectorÕs signal and to properly veto background events.
	
In the CDMS-DMC, prompt phonons are produced at the interaction point with 2~THz energy and isotropic distribution of wave momentum and slow-transverse, fast-transverse and longitudinal mode density of states weighted by $\langle v_p \rangle ^{-3}$ , where $v_p$ is the phase velocity and brackets indicate an average. Their propagation is described by quasidiffusion, and the CDMS-DMC incorporates phonon focusing, isotope scattering and an isotropic approximation for anharmonic decay. 
The phonon phase velocity given by the equation 
\begin{equation*}
\rho \omega^2 \epsilon_\mu = \sum_{\tau} \left(\sum_{\sigma \nu}
c_{\mu \sigma \nu \tau} k_\sigma k_\nu\right) \epsilon_\tau,
\end{equation*}
where $\rho$ is the crystal's mass density,\\
$\omega$ is the phonon frequency,\\
$\epsilon_\mu$ is a component of the polarization vector
$\mathbf{\epsilon}$,\\
$c_{\mu \sigma \nu \tau}$ is the elastic constant tensor and,\\
$k_{\sigma}$ is a component of the phase velocity vector
$\mathbf{k}$~\cite{Kittel1980}.

Group velocity is found by solving \begin{equation*}\vec{v}_g (\theta, \phi) = \frac{\partial \omega ( \theta, \phi ) }{\partial \vec{k}},\end{equation*} leading to a highly anisotropic distribution of phonons in position space~\cite{Wolfe1998}. A finely spaced lookup table of phase velocities, group velocities and polarization vectors is generated at the beginning of the CDMS-DMC.

Isotope scatting causes high-energy phonons to diffuse as they propagate with a bulk scattering rate $\Gamma_I = B \nu^4$, and individual scattering rate given by \begin{equation*}\gamma \sim \frac{ |\vec{e}_{\lambda} \cdot \vec{e}_{\lambda ^\prime}|^2} {\nu^3_{\lambda ^\prime}}~\cite{Wolfe1998}. \end{equation*}  Anharmonic decay dominates at early times with a scattering rate $\Gamma_A = A \nu^5$~\cite{Wolfe1998}. The decay rate constant $A$ is the literature is generally given averaged over all modes~\cite{Msall1997}, in the CDMS-DMC we apply anharmonic decay only to the longitudinal mode phonons~\cite{Tamura1993_2} (with appropriate scaling of $A$). After sufficient downconversion, the phonons can propagate ballistically to the detector surfaces.

At the detector surfaces, they either reflect or absorbed. Phonons absorbed on non-instrumented walls are removed from the MC. The description of absorption on instrumented surfaces is more detailed however. An energy of at least twice the aluminum superconducting gap must be deposited in the aluminum fins while the remainder could either remain in the film and be readout or it could be reintroduced into the crystal in a downconversion process. Phonons with energy below twice the aluminum superconducting gap cannot be readout in our phonon channels and are removed from the MC.

\section{The CDMS Detector Monte Carlo, Charge}

\begin{figure}
\begin{center}
\includegraphics[width=12cm]{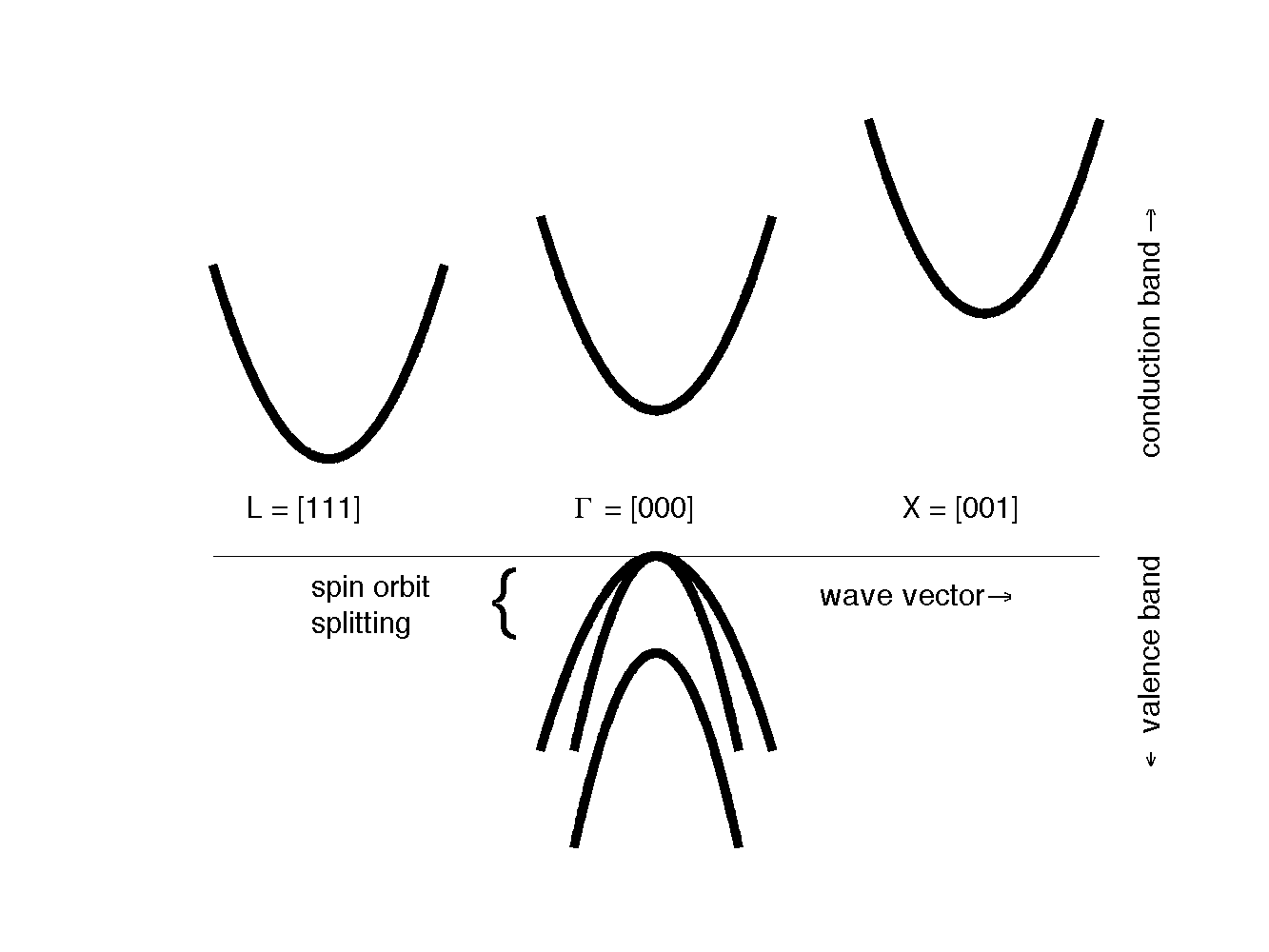}
\end{center}
\caption[] { \label{fig:bandGe} Germanium band structure showing the hole ground state, $\Gamma$ band and electron ground state L bands.}
\end{figure}

It is desirable to include a charge Monte Carlo for numerous reasons. First, the ionization signal, compared to the phonon signal, provides a discriminator between electron-recoil and nuclear-recoil events in the CDMS detectors. Electron transport is described by a mass tensor, leading to electron transport which is oblique to the applied field~\cite{Sasaki1958, Jacoboni1983} and is necessary to explain and interpret signals in the primary and guard-ring ionization channels~\cite{Cabrera}. Second, charges drifting through the detector produce a population of phonons, which contribute $\sim$25\% of the total phonon signal. These phononsÕ spatial, time, energy and emitted-direction distributions should therefore be properly modeled in the CDMS-DMC. Third, phonons created during electron-hole recombination at the surfaces contribute $\sim$25\% of the total phonon signal but in a low frequency, ballistic regime that is used to provide a surface-event discriminator. These phonons also need to be properly modeled in the CDMS-DMC. Germanium has an anisotropic band structure described schematically in Figure~\ref{fig:bandGe}. At low field, and low temperature, germaniumÕs energy band structure in such that the hole ground state is situated in the $\Gamma$ bandÕs [000] location and the electron ground state is in the L-band [111] location. Hole propagation dynamics are relatively simple due to propagation in the $\Gamma$ band and the isotropic energy-momentum dispersion relationship $\epsilon_{hole}(\mathbf{k}) = \hbar^2 k^2 / 2m$. Electron propagation dynamics are significantly more complicated due to the band structure and anisotropic energy-momentum relationship. At low fields and low temperatures, electrons are unable to reach sufficient energy to propagate in the $\Gamma$ or X-bands, and are not considered in the DMC. The electron energy-momentum dispersion relationship is anisotropic and given by $\epsilon_{electron}(\mathbf{k}) = \hbar^2 /2 (k_{\parallel}^2 / m_{\parallel} + k_{\perp}^2 / m_{\perp})$, where the longitudinal and transverse mass ratio $m_\parallel/m_\perp \sim$19.5.
  
\begin{figure}
\begin{center}
\includegraphics[width=12cm]{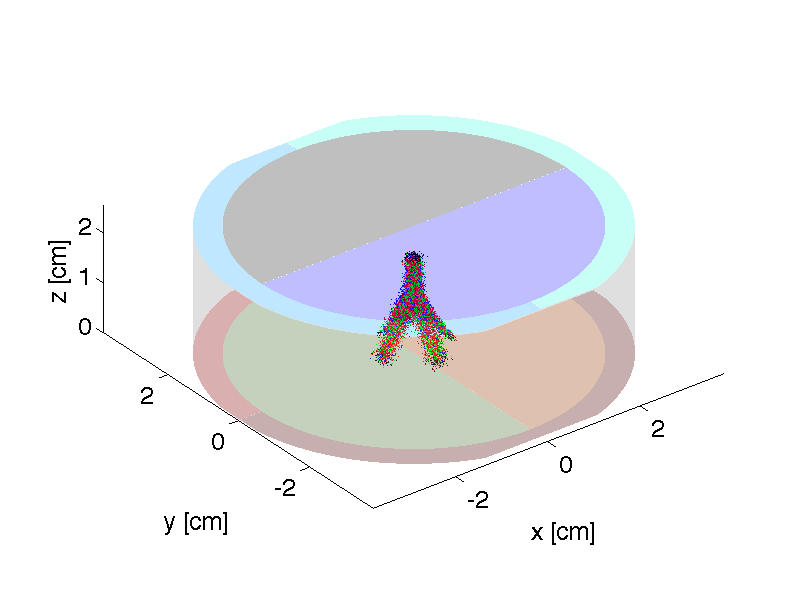}
\end{center}
\caption[] { \label{fig:129} Gamma-ray event in the CDMS-DMC. Holes propagate upwards in the $\Gamma$ band. The electrons propagate oblique to the electric field in the four L bands. Charge carriers are shown as black points. Slow transverse, fast transverse and longitudinal phonons are shown as red, green and blue points, respectively.}
\end{figure}

The SCDMS detectors are generally operated with low $\sim$1~V potential difference between the ionization channels and the phonon channels (which are held at ground). The potential accelerates and drift the charge carriers towards the appropriate detector face. As shown in Figure~\ref{fig:129}, the holes propagate parallel to the field, in the $+z$ direction. The electrons however, due to the band anisotropy, propagate in the four L valleys and oblique to the electric field ($\sim$33 degrees from the $-z$ axis).
		
\begin{figure}
\begin{tabular}{c}
\begin{minipage}{0.5\hsize}
\begin{center}
\includegraphics[width=7cm]{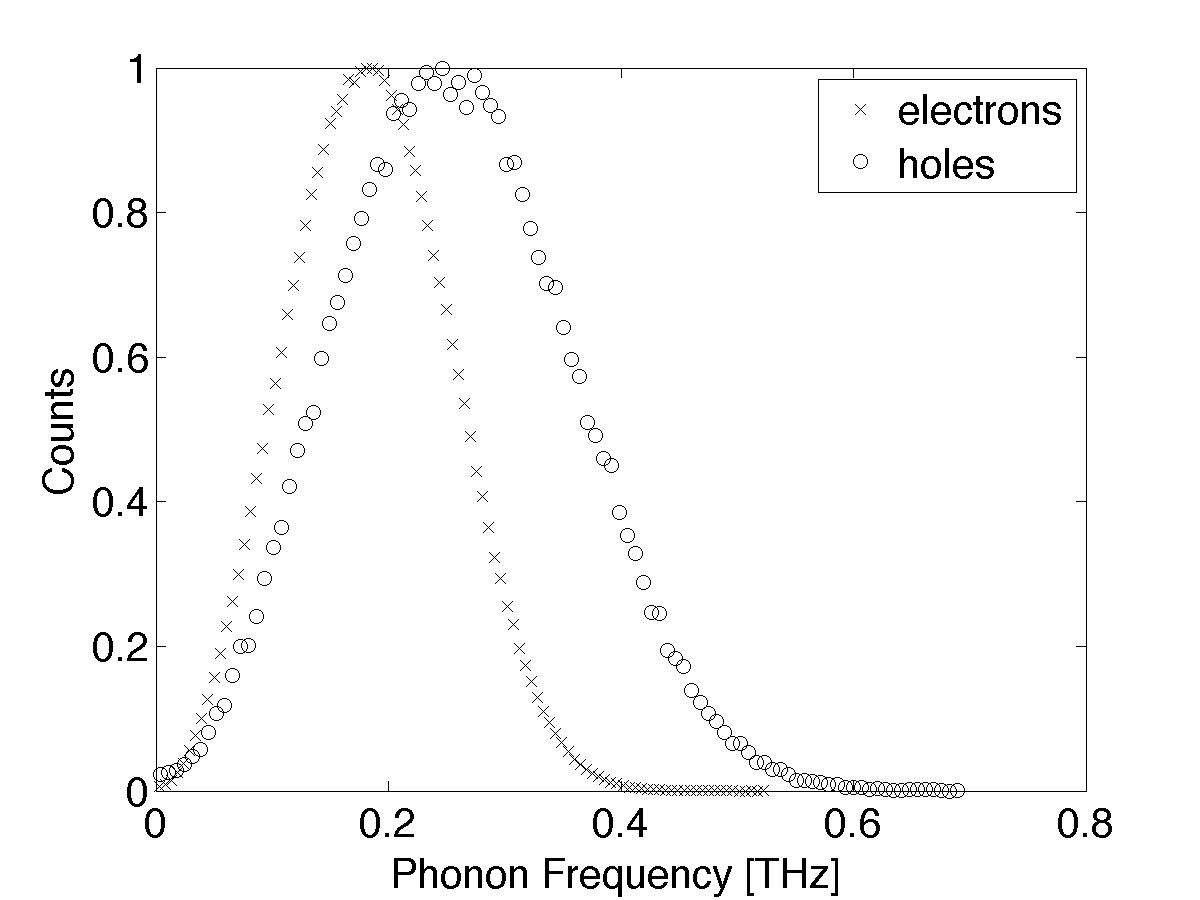}
\end{center}
\end{minipage}
\begin{minipage}{0.45\hsize}
\begin{center}
\includegraphics[width=7cm]{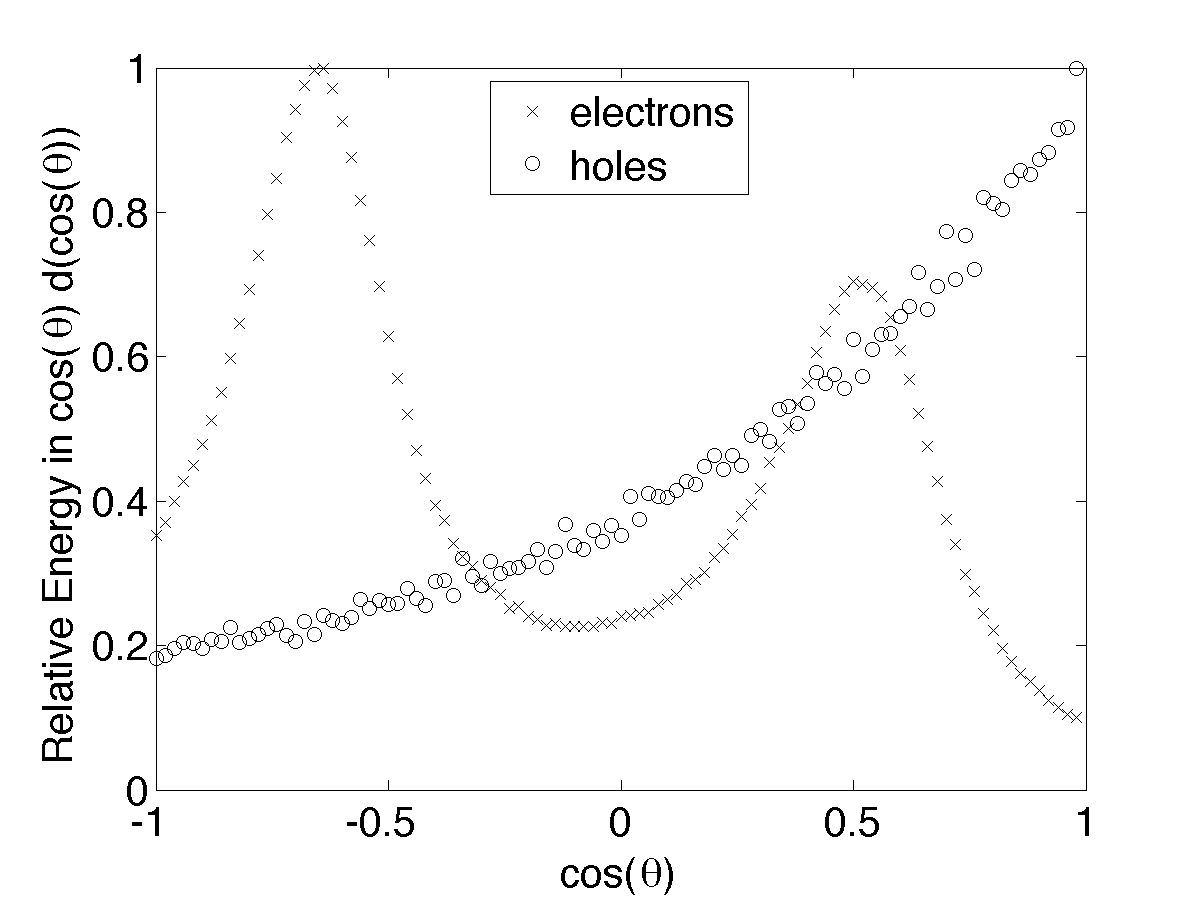}
\end{center}
\end{minipage}
\end{tabular}
\caption[] { \label{fig:1V_Nu}(left) Phonon energy spectra of electrons and holes drifted in [001] germanium with a 1~V field pointed in the [001] direction as determined in the CDMS-DMC. The spectra are normalized to peak heights of unity. Many of the carriers have frequency $<$0.3~THz frequency, and are ballistic.}
\caption[] { \label{fig:1V_cosTheta}(right) Angular distribution of emitted phonon energy relative to the crystalÕs [001] symmetry axis. In equilibrium, the net phonon momentum must be in the [001] direction, but there is a significant amount of energy back scattered into the direction opposite of the net drift velocity.}
\end{figure}

The charge carriers cannot accelerate indefinitely via the field and eventually scatter off of phonons to limit their velocity to the longitudinal phase velocity. Modeling of these scattering processes is complicated by anisotropy and can be handled either via a Herring-Vogt~\cite{Herring1956, Cabrera} transformation, or rejection sampling methods~\cite{Sundqvist2009}. At these low fields, the electrons (holes) emit phonons with a peak frequency of $\sim$0.2 (0.3)~THz (shown in figures~\ref{fig:1V_Nu} and~\ref{fig:1V_cosTheta}), and are ballistic.
 
\section{Experiments}

In the first round of experiments, a SCDMS mZip detector was exposed with an Am-241 gamma source to produce a population of prompt phonons and charge carriers within the detector. The detector ionization channels were biased at 6~V, producing a large population of Luke phonons. The detector was operated with the outer phonon channel segmented into three channels, which were individually instrumented and rotated 60~degrees relative to the inner three channels. In the data analysis, pulse starting time delays were used to determine position estimates, from which we selected events such that we obtained a spatially uniform distribution of calibration events. 
		
The version of the CDMS-DMC used in this analysis did not consider the oblique propagation of electrons and contained an analytical model for Luke phonon production, with longitudinal intravalley scattering. Holding all other quantities fixed, we varied the phonon absorption probability ($Abs_{phonon-side}$) on the side of the detector with instrumented phonon channels over the values (0.01, 0.03, 0.1, 0.15, 0.2 and 0.3). The side of the detector with instrumented ionization channels contained 23\% of the aluminum coverage as the phonon channel side and the phonon absorption probability was scaled accordingly. We also varied the probability of a phonon being absorbed on the non-instrumented walls ($Abs_{wall}$) from (0.001, 0.003, 0.01, 0.03, 0.1 and 0.3) for a total of 36 MC variations. Phonons were absorbed without downconversion on all surfaces.
		
\begin{figure}
\begin{tabular}{c}
\begin{minipage}{0.5\hsize}
\begin{center}
\includegraphics[width=7cm]{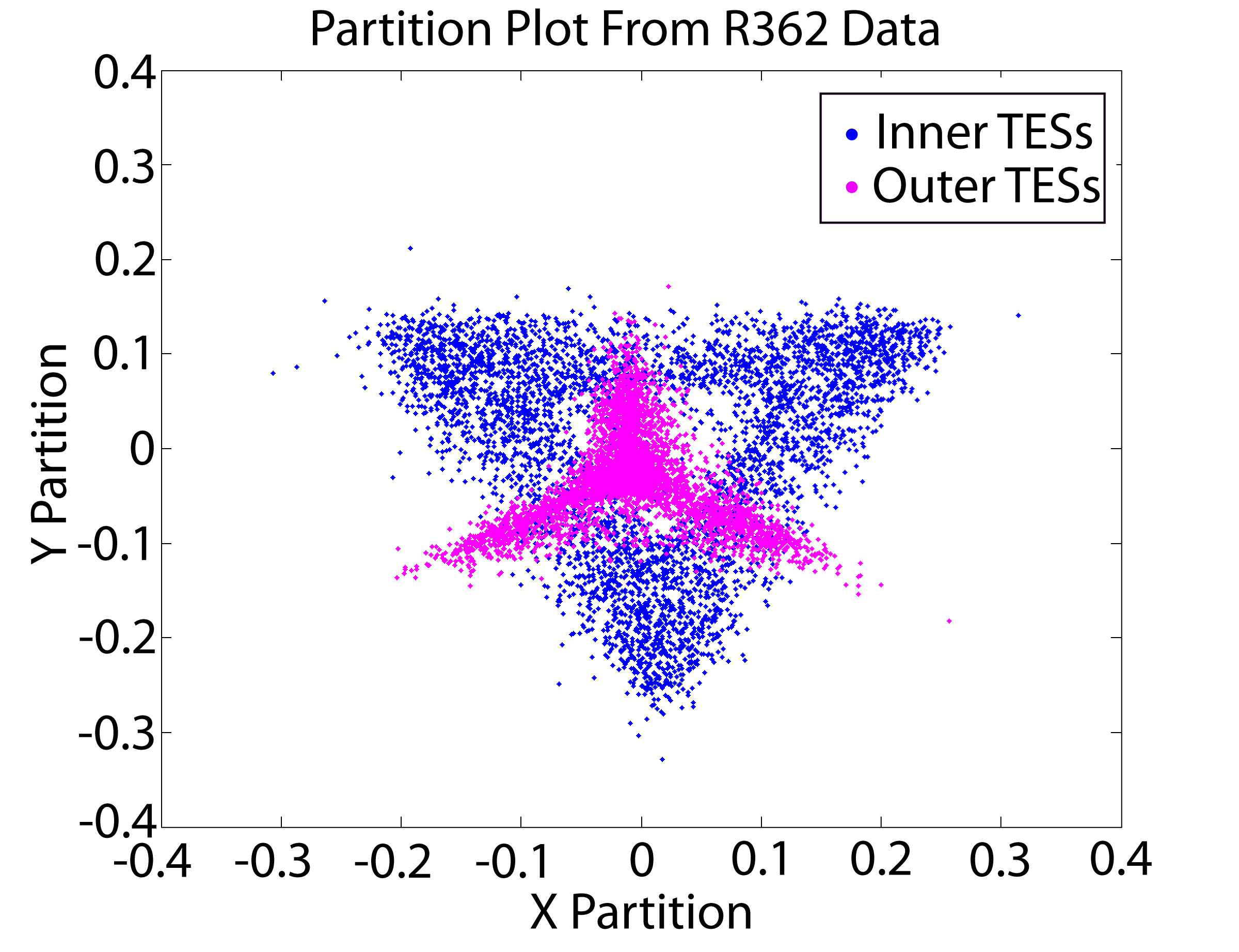}
\end{center}
\end{minipage}
\begin{minipage}{0.45\hsize}
\begin{center}
\includegraphics[width=7cm]{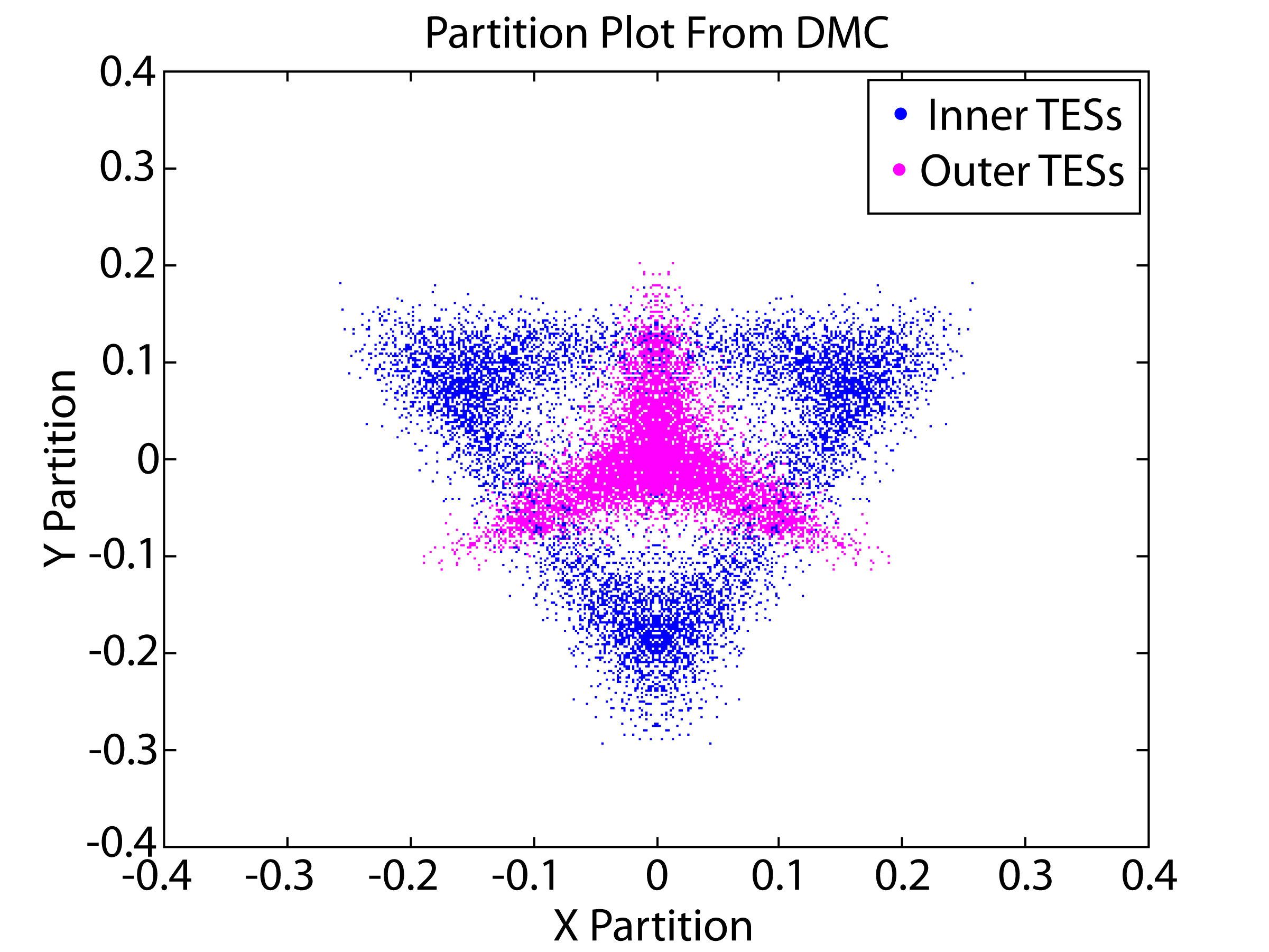}
\end{center}
\end{minipage}
\end{tabular}
\caption[] { \label{fig:fig6a}Event position estimates using the partitioning of phonon energy between phonon sensors. Two independent estimates are made with the three inner (triangle shaped partition plot) or three outer (crosshair shaped partition plot) phonon channels. (Left) event position in a CDMS mZip calibration run. (Right) CDMS-DMC results with ($Abs_{phonon-side}, Abs_{wall}$) = (0.1, 0.03).}
\end{figure}

The partitioning of phonon energy in the different phonon channels was used to determine a first moment position estimate. The inner three phonon channels and outer three phonon channels were considered separately to produce two position estimates for each event. Comparisons were made between the calibration data and CDMS-DMC results and models with ($Abs_{phonon-side}, Abs_{wall}$) = (0.1, 0.03), (0.15, 0.01) and (0.15, 0.03) providing good, qualitative agreement as shown in Figure~\ref{fig:fig6a}. The partition plots extent matches well, but the CDMS-DMC results are not as diffuse as the calibration data. 

\begin{figure}
\begin{tabular}{c}
\begin{minipage}{0.5\hsize}
\begin{center}
\includegraphics[width=7cm]{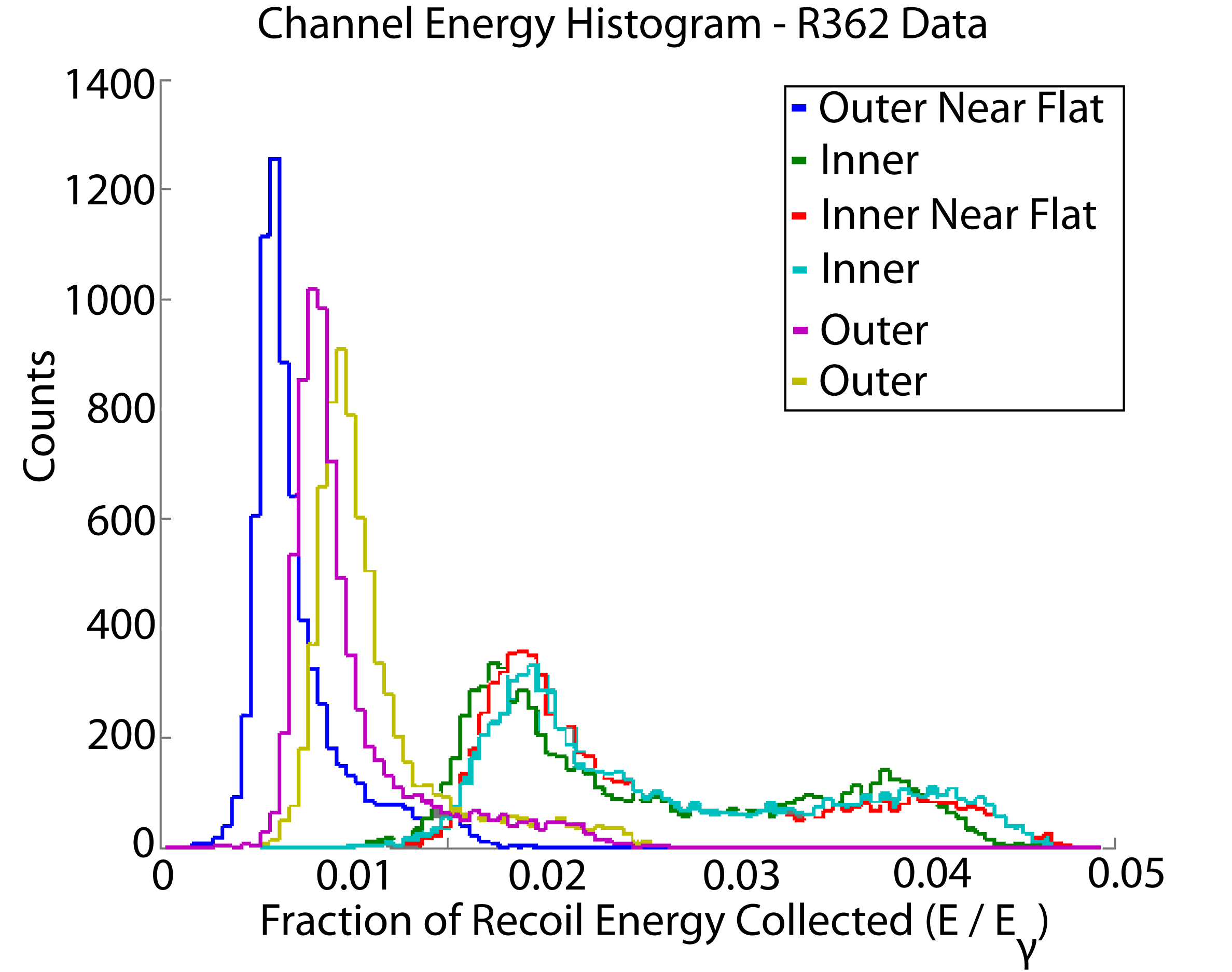}
\end{center}
\end{minipage}
\begin{minipage}{0.45\hsize}
\begin{center}
\includegraphics[width=7cm]{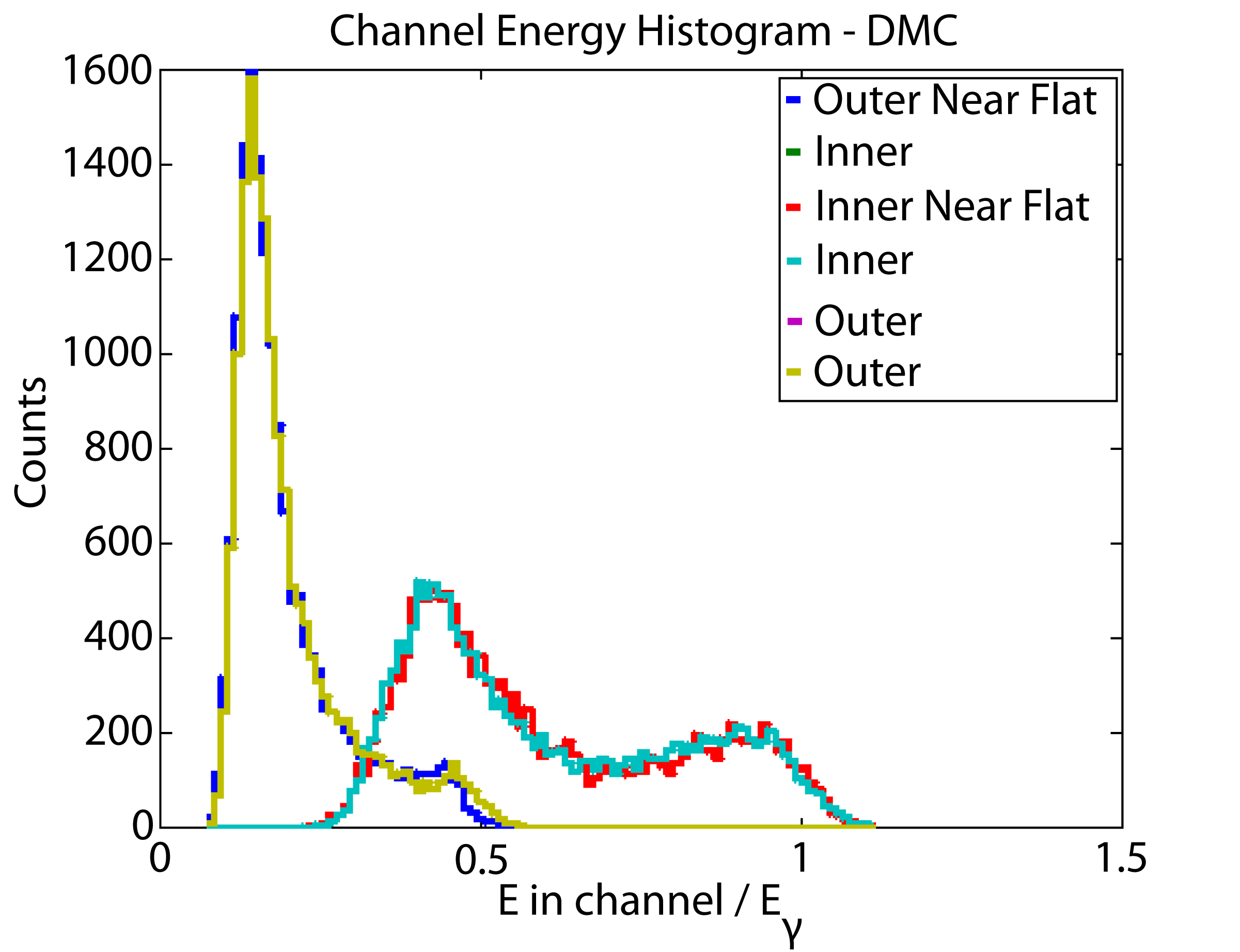}
\end{center}
\end{minipage}
\end{tabular}
\caption[] { \label{fig:fig7a}Histograms of phonon-flux energy deposited in the six phonon channels. (Left) CDMS mZip calibration data. (Right) CDMS-DMC results with ($Abs_{phonon-side}, Abs_{wall}$) = (0.15, 0.03).}
\end{figure}

Histograms of the energy absorbed in each of the six, phonon channels were produced for both the calibration data and CDMS-DMC results. Better qualitative agreement was seen for models with ($Abs_{phonon-side}, Abs_{wall}$) = (0.15, 0.01) and (0.15, 0.03) as shown in Figure~\ref{fig:fig7a}. The relative location of peaks and their heights are in good qualitative agreement.

Additional measurements were also made of the phonon decay time in both the calibration data and the CDMS-DMC results. The mZip calibration data showed pulse decay times of $\sim$300~$\mu$s. Parameters that were used in the CDMS-DMC and bracket the calibration data decay time include absorption probabilities of \\ ($Abs_{phonon-side}, Abs_{wall}$) = (0.03, 0.03), (0.1, 0.001), (0.1, 0.01) and (0.1, 0.03) with associated decay times of $\sim$400, 250, 250 and 200~$\mu$s respectively.

The iZip style CDMS detector also underwent a calibration run and comparisons were made between its measured phonon decay times and compared to the CDMS-DMC results. The reduced aluminum coverage, compared to the mZip, resulted in significantly longer pulse decay times of $\sim$900~$\mu$s. CDMS-DMC parameters that gave pulse decay times that bracketed the measured pulse decays included absorption probabilities of ($Abs_{phonon-side}, Abs_{wall}$) = (0.0013, 0.01), (0.0013, 0.03), (0.0039, 0.01), (0.013, 0.003), (0.013, 0.01) with associated decay times of $\sim$700, 800, 700, 900 and 700~$\mu$s. Based on the ratio of aluminum absorption coverage in the iZip and mZip detectors, an absorption probability of $Abs_{phonon-side,iZip}$ corresponds to $Abs_{phonon-side,mZip} / 7.7$. The iZip $Abs_{phonon-side,iZip}$ of 0.013 is consistent with the energy partitioning and phonon decay time measurements in the mZip style detector indicating good agreement between calibration data and the models with only the aluminum coverage area tuned down. The ratio of aluminum coverage to absorption probability indicates that, when incident on an aluminum fin, a phonon has a $\sim$21\% averaged probability of being absorbed.

\section{Conclusion}

The phonon quasidiffusion measurements presented here occurred on large 3~inch diameter, 1~inch thick high purity germanium crystals. These crystals were cooled to 50~mK in the vacuum of a dilution refrigerator. Phonon losses in the TESs and non-instrumented surfaces can were determined via the partitioning of energy and signal timing in the phonon channels. Good agreement is seen between the mZip and iZip CDMS-DMC models with only a corresponding adjustment in aluminum surface coverage. We conclude that there is a $\sim$21\% probability of phonons being absorbed when incident on aluminum. There was a smaller, 0.1\% chance of losing phonons on non-instrumented surfaces.

\section{Acknowledgements}

This research was funded in part by the Department of Energy (Grant Nos. DE-FG02-04ER41295 and DE- FG02-07ER41480) and by the National Science Foundation (Grant Nos. PHY-0542066, PHY-0503729, PHY-0503629, PHY-0504224, PHY-0705078, PHY-0801712)

\bibliographystyle{spiebib}
\bibliography{bibfile}

\end{document}